\documentclass[1p]{elsarticle}
\usepackage{subcaption}
\usepackage{graphicx}

\usepackage{amsmath}
\usepackage{amsfonts}
\usepackage{amssymb}

\usepackage{lineno}
\usepackage{hyperref}
\usepackage{color}

\journal{arXiv}

\biboptions{authoryear}

\begin{document}

\begin{frontmatter}

\title{Forecasting of Jump Arrivals in Stock Prices: New Attention-based Network Architecture using Limit Order Book Data}


\author[milla_moncef_address]{Milla M\"akinen}
\ead{milla.makinen@tut.fi}

\author[juho_adderess]{Juho Kanniainen\corref{mycorrespondingauthor}}
\cortext[mycorrespondingauthor]{Corresponding author}
\ead{juho.kanniainen@tut.fi}

\author[milla_moncef_address]{Moncef Gabbouj}
\ead{moncef.gabbouj@tut.fi}

\author[alexandros_address]{Alexandros Iosifidis}
\ead{alexandros.iosifidis@eng.au.dk}

\address[milla_moncef_address]{ Laboratory of Signal Processing, Tampere University of Technology, Finland.}
\address[juho_adderess]{Laboratory of Industrial and Information Management, Tampere University of Technology, Finland.}
\address[alexandros_address]{Department of Engineering, Electrical and Computer Engineering, Aarhus University, Denmark}

\begin{abstract}
The existing literature provides evidence that limit order book data can be used to predict short-term price movements in stock markets. This paper proposes a new neural network architecture for predicting return jump arrivals in equity markets with high-frequency limit order book data. This new architecture, based on Convolutional Long Short-Term Memory with Attention, is introduced to apply time series representation learning with memory and to focus the prediction attention on the most important features to improve performance. The data set consists of order book data on five liquid U.S. stocks. The use of the attention mechanism makes it possible to analyze the importance of the inclusion limit order book data and other input variables. By using this mechanism, we provide evidence that the use of limit order book data was found to improve the performance of the proposed model in jump prediction, either clearly or marginally, depending on the underlying stock. This suggests that path-dependence in limit order book markets is a stock specific feature. Moreover, we find that the proposed approach with an attention mechanism outperforms the multi-layer perceptron network as well as the convolutional neural network and Long Short-Term memory model. 
\end{abstract}


\end{frontmatter}

\section{Introduction}
Nowadays, many exchanges, such as the New York Stock Exchange (NYSE) and various NASDAQ exchanges, are using systems driven by limit order submissions. Limit orders are submissions to the system that contain a price and the desired quantity to buy or sell. The Limit Order Book (LOB) markets operate in very high frequencies, where delays generally range from milliseconds to several nanoseconds for machines located near the exchange. This, along with the possibility to obtain event data from exchanges, yields huge amounts of data, which has created new opportunities for data processing. This enables market analysis on a completely new level on many interesting questions \cite[see, for example][]{toth2015equity,chiarella2015learning}, but has also brought unique challenges for both theory and computational methods \citep{Cont2011a}. In the recent literature, both tractable models and data-driven approach---that is, machine learning---have been introduced to predict price movements with LOB data \citep{Cont2010,Cont2011a,cont2012order,Kercheval2015a,ntakaris2017benchmark,tsantekidis2017using,tsantekidis2017forecasting,passalis2017time,dixon2018high,tran2018temporal,sirignano2018universal}. Overall, the existing literature provides evidence that limit order book data can be used to predict price movements in stock markets.

Even though stock prices movements have been predicted using LOB  data in general, less research is published about the use of LOB data to predict the arrival of {\em jumps} in stock prices. Stock price jumps are significant discontinuities in the price path so that realized return at that time is much greater than usual continuous innovations. In the literature, there is strong empirical evidence on the existence of return jumps  in stock markets \citep[see, e.g.,][and references therein]{eraker2004stock,Lee2012,yang2017jump}. Economically, the return jumps reflect information arrivals \citep{Lee2012,bradley2014analysts,kanniainen2017arrival}, and therefore the jumps in stock prices are also related to the predictability of information releases. Moreover, return 
jumps are fundamentally important in option pricing \citep{cont2003financial}. 

The main research question that this work addresses is as follows: How well can the arrival of jumps in equity returns be predicted using high-frequency limit order book (LOB) data with advanced machine learning techniques? The consequent question is if price jumps can be foreseen in the order book data. These questions are motivated by the fact that market makers, i.e. liquidity providers, can have prior information about forthcoming---scheduled or non-scheduled---news arrivals that will be realized as large price movements and play against the market makers. Sophisticated market makers do not want to provide liquidity because  limit orders can be understood as options to trade the underlying security at a given price and they suffer from adverse selection \citep[see][]{copeland83}. As \citet{foucault07} argue, speculators may exercise these options, that is, pick off limit orders, if limit orders become stale after the arrival of new information. For this reason, sophisticated market makers do not want to take a risk that their limit orders are on the wrong side of the book to be exploited by fast traders right after the price jumps, which is seen as low limit order book liquidity \cite{Siikanen2017}. Moreover, if  market makers were capable to predict not only the location but also the direction of the forthcoming jump, then limit order book can become asymmetrically illiquid. 

\begin{figure}[!ht]
  \begin{center}
    \includegraphics[width=1.1\textwidth]{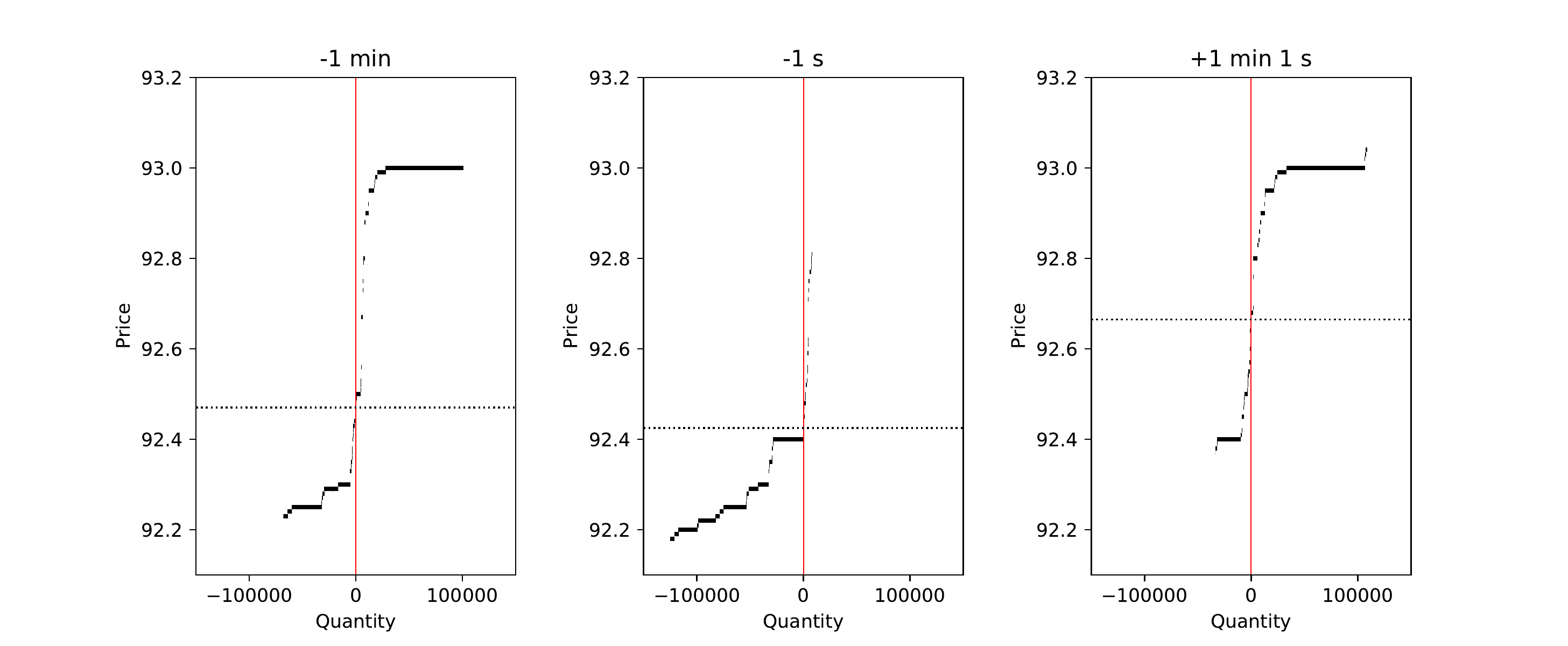}
  \end{center}
  \caption{Three snapshots on Apple's order book on 9th of June, 2014, around a jump detected between 9:33-9:34am. A plot on left shows the state of the order book one minute before the beginning of the jump interval at  9:32am. A plot in the middle shows the state of the order book just a second before the beginning of the interval at 9:29:59am. The third plot on right draws the order book a second after the end of the interval at 9:34:01am. The side left of the red line at zero quantity contains the bids (i.e. bid orders are presented as negative quantity), referred to as the bid side of the book, and the right side contains asks (i.e. positive quantity), referred to as the ask side. The black dotted lines present the mid-prices. The data is provided by Nasdaq US.
}
  \label{fig:appleJump}
\end{figure}

This kind of situation is demonstrated in Figure \ref{fig:appleJump}, which illustrates the order book states around a jump in mid-price for Apple on 9th of June, 2014, where a positive jump was detected between 9:33-9:34am. The figure provides snapshots for 1 minute and 1 second before the beginning of the 1-minute jump interval and the third snapshot plot 1 second after the end of the same 1-minute jump interval. It demonstrates the following:
\begin{itemize}
\item [--] 1 minute before the beginning of the jump interval: The order book is rather asymmetric, though quite thin (thin and widespread) and it is relatively expensive to trade a large number of shares by market order.
\item [--] 1 second before the beginning of the the jump interval: Order book has become asymmetric so that the order book is very illiquid on ask side while remaining relatively liquid on bid side. This can mean that liquidity providers had a hunch on shortly arriving positive mid-price jump. In this case, even small trades on ask side can induce large price movements up.
\item [--] 1 second after the end of jump interval (and 1 minute 1 second after the beginning of the interval): The liquidity provides have come back on ask side and the ask-side liquidity has recovered.

\end{itemize}

This example also begs the question about the root cause: if market makers anticipated the price jump based on market fundamentals and thus delivered no liquidity on ask side or, alternatively, if the price movement was introduced by microstructure noise so that the order book illiquidity on the ask side was not based on market fundamentals. In this paper, we keep both explanation possible. In fact, the root cause is rather irrelevant -- the aim of this paper is to build neural network models to predict price jumps and thus answer the question if price jumps can be foreseen in the order book data, were the root cause one or other.  

Methodologically, we have two-class prediction problem: Whether or not there is a jump within the next minute. The output data consists of minute-by-minute observations about the location and sign of detected jumps in stock prices. The input data is extracted from the reconstructed order books. We do not only use the 'raw' data, i.e. prices and quantities on different levels, but also hand-crafted features that are extracted from millisecond-level observations over the past 120 minutes. Regarding jumps, this paper follows the existing literature to define jumps as large price movements that cannot be explained by Brownian motion. As a preliminary step, the locations of return jumps are detected from the high-frequency mid-price data utilizing the nonparametric jump detection test of \citep{Lee2007}. Then, after preprocessing the limit order book data, various neural network methods are applied to predict the locations of jumps using real-time features on high-frequency limit order data.

In this paper, machine learning refers to a group of methods characterized by their learning property, which allows the system to adjust its parameters by itself. Different machine learning methods, especially neural networks, the first of which introduced in the 1950s \citep{Rosenblatt1957}, have been becoming increasingly popular within the last decade. Neural networks have been shown to be one of the few methods that is broadly successful in time series prediction \citep{Graves2012}, although financial time series are generally regarded as very difficult to predict \citep{Kara2011}. In this paper, we use not only the standard multi-layer perceptron network (MLP) but also a convolutional neural network (CNN) and a Long Short-Term Memory (LSTM) network, both of which have been especially successful in predicting stock price movements \citep{tsantekidis2017using,tsantekidis2017forecasting}. Moreover, a new network model is developed by combining convolutional and Long Short-Term Memory (LSTM) layers as well as the attention model proposed by \citet{Zhou2016}. The proposed convolutional LSTM attention model (CNN--LSTM--Attention) aims to utilize LSTM for time series memory, convolution (CNN), and the attention model for reducing the input size, increasing locality, and focusing on the most important features to improve prediction results. In addition to the main question above, we also consider which method (MLP, CNN, LSTM, CNN-LSTM-Attention) is best for predicting jumps with LOB data. The performance of the proposed CNN-LSTM-Attention network is of particular interest, as it offers a new combination of methods that is jointly optimized for jump prediction.

To analyze the predictability and performance of the selected networks, a dataset of high frequency LOB data from several top NASDAQ stocks is employed for both training and testing the proposed methods. The stocks that are used are GOOG (Google), MSFT (Microsoft), AAPL (Apple), INTC (Intel), and FB (Facebook).\footnote{We emphasize that the proposed methods {\em are applicable for  any} security for which limit order book data is available. At the same time, the methods {\em are not applicable} to predict jumps in Foreign Exchange Markets \citep{bates1996jumps} and other markets where such a limit order book is not publicly available or to analyze the processes related to the real investments \citep{dixit1994investment,kanniainen2009can} or other assets whose value processes are not observable.}

The rest of the paper is organized as follows. Section \ref{SEC:Data} introduces both output (detected jump locations) and input data (real-time order book features). Then, Section \ref{SEC:NeuralNetworks} presents the network models used in this paper, including the new network architecture called the CNN-LSTM-Attention model. Section \ref{SEC:Results} provides the empirical results, and, finally, Section \ref{SEC:Conclusions} concludes this work.

\section{Data}\label{SEC:Data}

\subsection{Data sets}

This research is conducted using NASDAQ's ``TotalView-ITCH'' limit order data. The data consist of ultra-high-frequency (on millisecond bases) information regarding the limit orders, cancellations, and trades executed through NASDAQ's system. The data contain prices and quantities of orders as well as their linked partial and full trades and cancellations. The data are further transformed into two data sets:
\begin{itemize}
\item [(i)] Output data: Minute-by-minute data about detected jumps in stock prices. This is based on mid-price observations from which jumps are detected, so each one-minute time period is classified as either having a jump or not.
\item [(ii)] Event-by-event input data about the state of the order book. These data are extracted from millisecond-level observations about order book events. The resulting data contain both bid and ask prices as well as their quantities for the ten best levels on both sides of the book.
\end{itemize}

The order-driven market systems work in a way such that investors may place either ask or bid orders of their desired price, and the system will match eligible orders to create a trade. Orders may be either limit or market orders. Limit orders are placed in the list of orders at a specified price. Market orders are immediately executed with the limit order of the best price if such exists. In a way, this resembles a queue system, especially when orders of identical prices are submitted. A limit order that has not been executed can also be cancelled at any time. Both trades and cancels can also be partial, meaning that a part of the limit order will be left in the book after execution \citep{Cont2010}.

\begin{table}[h]
\centering
\bgroup
\def\arraystretch{1.2}
 \begin{tabular}{|c|c|c|c|}
 \hline
Stock & Orders & Trades & Cancels \\
 \hline
AAPL & 1963.37 & 181.33 & 1870.52 \\
FB & 1665.53 & 136.32 & 1563.80 \\
INTC & 848.58 & 71.38 & 823.11 \\
MSFT & 1304.75 & 95.22 & 1272.25 \\
GOOG & 480.34 & 27.86 & 462.20 \\
 \hline

\end{tabular}
\egroup
\caption{Average number of order submissions, trades, and cancels for each stock over a minute. Trades and cancels also include partial executions and cancellations of orders.}%
\label{table:average_over_minute}
\end{table}

To ensure a continuous order flow, several well-known liquid stocks are selected for the study. These are GOOG (Google), MSFT (Microsoft), AAPL (Apple)\footnote{The price data for AAPL was adjusted slightly. On June 6, 2014 at 5pm, Apple issued 10 800 000 000 new shares, effectively splitting each existing common share into seven separate parts. As this was in the middle of the observed period and caused no difference in individual investors' wealth in terms of owned stock, all stock prices prior to the split are divided by seven to make the true value of the owned stock continuous.}, INTC (Intel), and FB (Facebook). All of the selected stocks have large amounts of orders and trades each day. Table \ref{table:average_over_minute} shows the average numbers of order submissions, cancellations, and trades over one minute.

The data are divided into two categories: training data and test data. The training data are those used to learn the problem, that is, the data fed to the networks in the training phase to adjust the weights through the optimization algorithm. The training data consist of the series of observations in fifty-days periods. Fifteen percent of the training data is selected as validation data before starting the training of a model. Validation and test data are intended to evaluate the performance of the system. This cannot be done with the training data alone, as the model will easily be overfitted, which sharply reduces the performance outside the training dataset because the trained model is no longer generalizable \citep{Webb2011}. The difference between test and validation data is that validation data is constantly used in model selection and adjustment during the training phase. After selecting the best model, test data are used to evaluate the model's performance. Thus, validation data are kept separately from test data to ensure that the model is not developed solely to be able to classify the test data; moreover, this provides an objective view on the performance of the system.

In all datasets, observations are picked every minute, but the amount of jump samples is increased by duplicating the jump observations. Specifically, the beginning of the duplicated sample is shifted by several seconds to ensure there are no identical samples.  The time intervals of the data sets are presented in Table \ref{table:training_and_test_data}. Validation data are selected in a way such that the duplicated samples belong to either the training or validation data sets. The data are divided into training sets based on the day of the observation. A total of 360 days, spanning about one and half years, are selected from 2014-2015. The data are divided so that first there are 50 days of training data, followed by 10 days of test data. The next set contains the first 50 days as well as the following 50, and it is tested on the following 10 days after both sets. This pattern is followed through the whole dataset so that the seventh test set trains on 350 days and tests with the last 10 of 360. Additionally, the training data are presented in a window such that the model is trained on the newest 50 samples at a time starting from the beginning of the observation period (but not reset between sets).

\begin{table}[h]
\centering
\bgroup
\def\arraystretch{1.2}
 \begin{tabular}{|c|c|c|}
 \hline
 Data set & Training days & Test days \\
 \hline
 1. & 1-50 & 51-60 \\
 2. & 1-100 & 101-110 \\
 3. & 1-150 & 151-160 \\
 4. & 1-200 & 201-210 \\
 5. & 1-250 & 251-260 \\
 6. & 1-300 & 301-310 \\
 7. & 1-350 & 351-360 \\
 \hline

\end{tabular}
\egroup
\caption{Division of data into sets used in training in 50 daylong sequences.}%
\label{table:training_and_test_data}\end{table}

\subsection{Detected Jumps (output data)}
To detect jumps in stock prices, we use an algorithm proposed by \citet{Lee2007}. As jumps are predicted short term, samples are collected every minute for the duration of the observation period. This gives a one-minute window in which a jump may occur, allowing these samples to be classified as either having or not having a jump in the following one-minute period. We run the jump detection algorithm for the entire sampling period for the collection of necessary amounts of jump samples. The length of the data window used for the estimation of bipower variation is 600 minutes.

\begin{figure}[!ht]
  \begin{center}
    \includegraphics[width=0.6\textwidth]{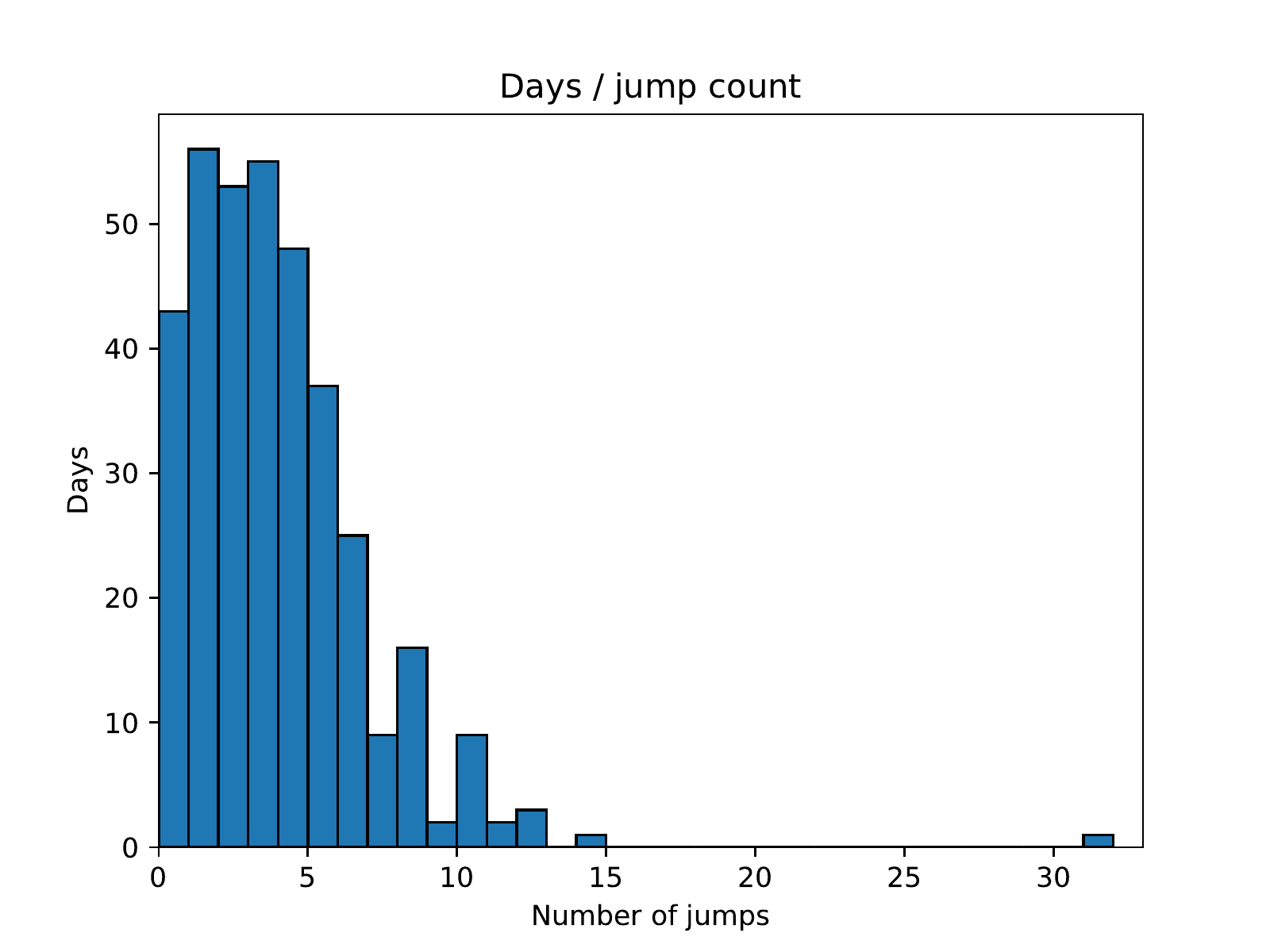}
  \end{center}
  \caption{Jumps per day counts for AAPL. Around 12\% of days had no jumps, and around 19\% of
days had more than five jumps, with the median being three jumps.}
  \label{fig:j_day}
\end{figure}

\begin{figure}[!ht]
  \begin{center}
    \includegraphics[width=0.6\textwidth]{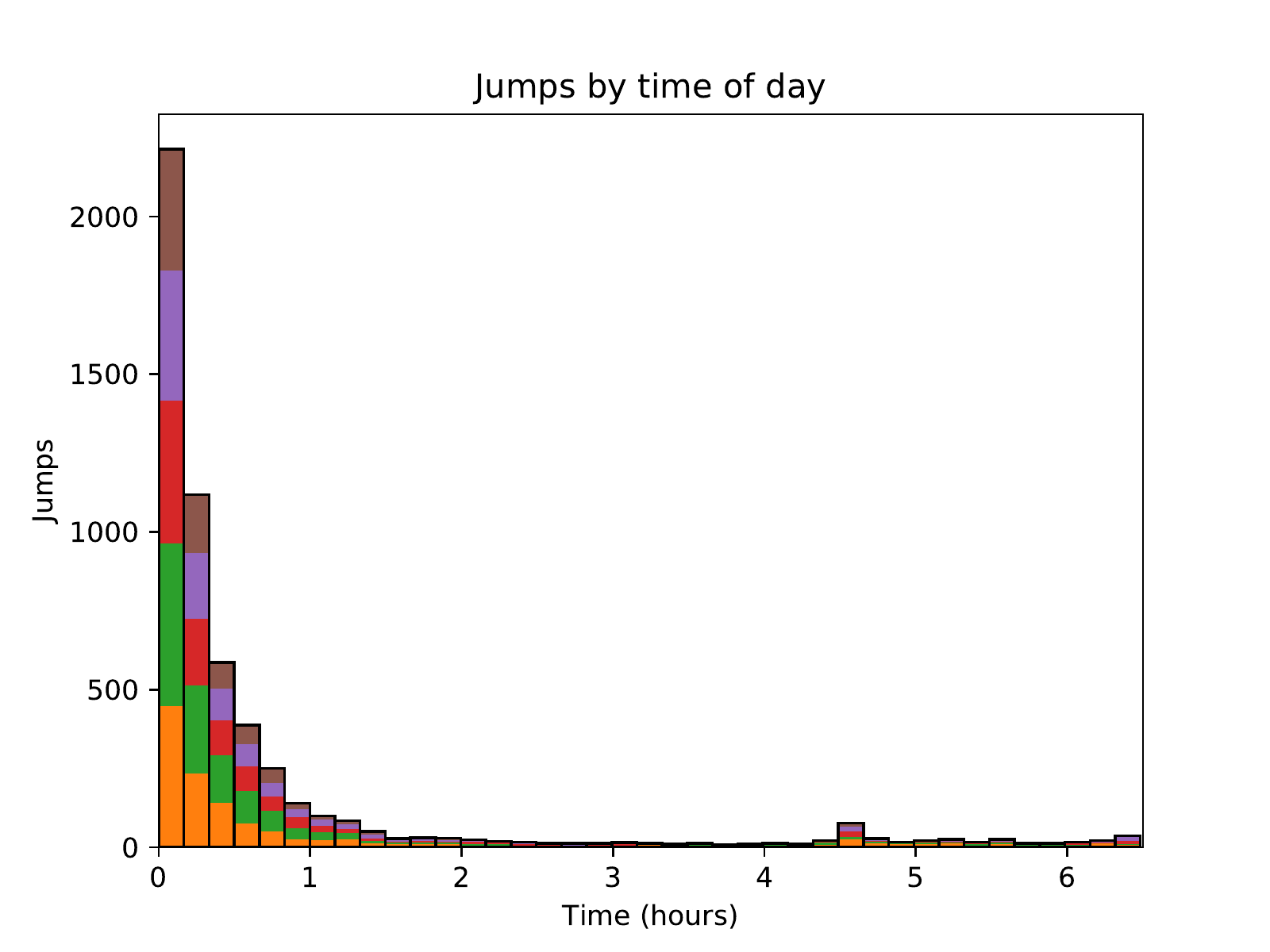}
  \end{center}
  \caption{Total amount of jumps at times of the day in 10 minute periods starting from the beginning of the trading day at 9.30 am. All stocks are distributed similarly, shown by the different colors, from top to bottom: AAPL, FB, INTC, MSFT, GOOG.}
  \label{fig:j_day_2}
\end{figure}

The frequencies of detected jumps are presented in Table \ref{table:freq_of_jumps}. On average, there are around three jumps per day per stock. However, jumps are not evenly divided between days. Instead, the days that have jumps tend to have a larger number of jumps on average. A sample distribution of jumps per day counts is shown in Figure \ref{fig:j_day}. Moreover, during a single trading day, jumps tend to be heavily skewed towards morning hours, as observed, for example, by \citet{Lee2007}. The vast majority of detected jumps occurred within the first half hour of the trading day, with only occasional jumps after the first 1.5 hours for all stocks. Additionally, all stocks had a slight increase in quantity at 2 pm, where the time period between 14:00 and 14:05 contained around four times as many jumps as between 13:55 and 14:00. The jumps at this time occur during multiple days thorough the whole observation period. The distribution of jumps according to the time of day counted for the whole observation period is presented in Figure \ref{fig:j_day_2}.

\begin{table}[h]
\centering
\bgroup
\def\arraystretch{1.2}
\begin{tabular}{|c|ccccc|c|}
\hline
Training period & AAPL & FB & GOOG & MSFT & INTC & Average \\
\hline
1-50 & 164 & 182 & 155 & 160 & 149 & 162 \\
51-100 & 200 & 177 & 131 & 161 & 150 & 164 \\
101-150 & 172 & 192 & 102 & 152 & 165 & 157 \\
151-200 & 161 & 171 & 125 & 132 & 170 & 152 \\
201-250 & 178 & 181 & 136 & 149 & 155 & 160 \\
251-300 & 172 & 186 & 111 & 128 & 155 & 150 \\
301-350 & 184 & 182 & 109 & 122 & 139 & 147 \\
\hline
Average & 176 & 182 & 124 & 143 & 155 & 156 \\
\hline
\end{tabular}

\vspace{0.5cm}

\begin{tabular}{|c|ccccc|c|}
\hline
Test period & AAPL & FB & GOOG & MSFT & INTC & Average \\
\hline
51-60 & 37 & 35 & 42 & 37 & 38 & 38 \\
101-110 & 54 & 55 & 37 & 27 & 17 & 38 \\
151-160 & 49 & 38 & 38 & 26 & 27 & 36 \\
201-210 & 32 & 41 & 26 & 28 & 18 & 29 \\
251-260 & 39 & 32 & 33 & 26 & 27 & 31 \\
301-310 & 35 & 35 & 12 & 24 & 18 & 25 \\
351-360 & 26 & 19 & 16 & 5 & 13 & 16 \\
\hline
Average & 39 & 36 & 29 & 25 & 23 & 30 \\
\hline
\end{tabular}

\egroup

\caption[Frequencies of jumps in the dataset]{The frequencies of jumps in
the training and test datasets by stock and set. A total of 5537 jumps across $362
- 2$ days were observed.}
\label{table:freq_of_jumps}
\end{table}

The data from the stocks are used to construct training and test sets by time and stock, as presented in Table \ref{table:freq_of_jumps}. Jumps at the very first observation of the public market opening (9.30) were not taken into account. Additionally, jumps from the first two days were not detected due to the insufficient amount of previous observations to satisfy the window size requirement of the jump detection algorithm. This also means that the training sets presented in Table \ref{table:training_and_test_data} skip the first two days of the sequence to avoid labeling possible jump samples as non-jumps due to the undetectability of jumps during the beginning of the price sequence. Thus, day 1 in the table is really day 3 of the price observation period.

\subsection{Order book state data (input data)}
The inputs use LOB data, which is reconstructed from the order book event data. The LOB contains both ask and bid prices as well as their quantities for the ten best levels on both sides of the book. This is done simply by checking active orders at a certain time, which can be then ordered by price to obtain the ten best levels so that the lowest ask and the highest bid are on the first level, and subsequent levels are filled by existing prices next in the order. The quantity is the sum of quantities for orders of that price, and quantities at levels with multiple orders are the sum of all active orders at that level. The method of constructing the book also means that empty levels cannot exist between two defined prices. Instead, completely empty ticks are left off unless there simply are not enough orders to fill the ten levels, in which case the levels last in order are filled with prices and quantities of 0.

To get the best view of the state of the order book, we follow \citet{Kercheval2015a} to extract 144 indicators from the data: a) the basic set of features containing the raw LOB data over ten levels, with both sides containing price and volume values for bid and ask orders, b) the time-insensitive set of features describing the state of the LOB, exploiting past information, and c) the time-sensitive features describing the information edge in the raw data by taking time into account. The time-insensitive set contains further information about the spreads, differences, and means. The time-sensitive set contains features that indicate changes in the data in time, such as derivatives, accelerations, and intensities. These features, provided in Table \ref {table:Features}, are used also in \citet{ntakaris2017benchmark,tsantekidis2017using,tsantekidis2017forecasting,passalis2017time,tran2018temporal}.

\begin{table}[ht]
\centering
\scalebox{0.85}{
\begin{tabular}{lll}
Feature Set & Description & Details   \\
\hline
a) Basic 				&	$ v_1 = \{ P_i^{ask}, V_i^{ask}, P_i^{bid}, V_i^{bid}\}_{i=1}^n$  											&	10-level LOB Data, $i=1\dots n$\\
\\
b) Time-		& 	$ v_2 = \{(P_i^{ask}-P_i^{bid}), (P_i^{ask}+P_i^{bid})/2 \}_{i=1}^n $										&	Spread \& Mid-Price\\
Insensitive					& 	$ v_3 = \{|P_{i+1}^{ask} - P_i^{ask}|, |P_{i+1}^{bid} - P_i^{bid}|\}_{i=1}^{n-1}$	&	Price differences\\
					& 	$ v_4 = \{ \frac{1}{n}\sum\limits_{i=1}^{n}P_i^{ask},  \frac{1}{n}\sum\limits_{i=1}^{n}P_i^{bid},  \frac{1}{n}\sum\limits_{i=1}^{n}V_i^{ask},  \frac{1}{n}\sum\limits_{i=1}^{n}V_i^{bid}\}$		&	Price \& Volume means	\\

					& 	$ v_5 = \{\frac{1}{n}\sum^n_{i=1}(P_i^{ask} - P_i^{bid}), \frac{1}{n}\sum^n_{i=1}(V_i^{ask} - V_i^{bid})\} $			&	Accumulated differences			\\
\\

c) Time-  		&	$v_6 = \{dP_i^{ask}/{dt}, dP_i^{bid}/{dt}, dV_i^{ask}/{dt}, dV_i^{bid}/{dt} \}_{i=1}^n $		&	Price \& Volume derivatives \\
Sensitive			&	$v_7 =  \{ \lambda^{la}_{\Delta t}, \lambda^{lb}_{\Delta t}, \lambda^{ma}_{\Delta t}, \lambda^{mb}_{\Delta t}, \lambda^{ca}_{\Delta t}, \lambda^{cb}_{\Delta t} \}$		&	Average intensity per type			\\
					&	$v_8 = \{ 1_{\lambda^{la}_{\Delta t} > \lambda^{la}_{\Delta T}}, 1_{\lambda^{lb}_{\Delta t} > \lambda^{lb}_{\Delta T}}, 1_{\lambda^{ma}_{\Delta t} > \lambda^{ma}_{\Delta T}}, 1_{\lambda^{mb}_{\Delta t} > \lambda^{mb}_{\Delta T}} \}$ 	&	Relative intensity indicators 			\\
					&	$v_9 = \{d\lambda^{ma}/dt, d\lambda^{lb}/dt, d\lambda^{mb}/dt, d\lambda^{la}/dt  \}$  	&	Accelarations 			\\

d) Clock time 		&	$v_{10} = \{\lfloor \frac{t}{60} \rfloor  \}$ &	Time, rounded to hours\\
\hline
\end{tabular}
}
\caption{Feature Sets. In the table, $P$ stands for prices and $V$ for volumes. In addition, $\lambda$ denoted the intensity of a given order book event.}\label{table:Features}
\end{table}

In addition to the LOB data, some of the time-sensitive features presented in \citet{Kercheval2015a} require calculating intensities, that is, the number of arriving orders or cancellations of a certain type, which cannot be directly calculated from the constructed book and instead must be counted from the original event data. The intensities are separated into ask and bid, and the orders are categorized based on whether they are limit or market orders. The intensities at each step are calculated directly from the order flow data and attached to the corresponding order book data of the step. Within market hours, both the limit order book state and the intensities are calculated every second, yielding a total of 23,400 observations per day. Data from non-trading hours are discarded due to different trading mechanisms, and the data used over multiple days is treated as a continuous sequence. In addition to some of the suggested features, approximate times of the observations are included to account for the differences in stock behavior at different points during the day. The timestamps are rounded to the nearest hour to avoid converging to the local minima of purely time-based classification.

For the data sets (Table \ref{table:training_and_test_data}), samples are extracted by a one-minute moving window thorough the training set, creating one sample per minute, for a total of 390 samples a day. Positive samples are defined as those with a jump right after the last observation, that is, during the next minute, which is not included in the window. Negative samples are only collected from the moving window; for positive samples, the window is shifted slightly multiple times to generate more positive samples due to the large difference in the sample sizes. As the data are collected every second, it is possible to shift the window small enough amounts to not include the jump while creating slightly different data for the samples to increase variety and to preserve the original classification of a jump existing within the next minute. To ensure that possible periodical changes in the order books will not affect the classification results due to only positive samples being shifted, negative samples are also shifted randomly.

All collected samples contain 120 steps sampled at a one-minute interval. These samples are then normalized using the z-score to eliminate the irrelevant noise due to, for example, different starting prices: $\mathbf{x}_{normalized} = \left(\mathbf{x} - \bar{x}\right)/{\sigma_{\bar{x}}}$, where $\mathbf{x}$ is the feature vector to be normalized, $\bar{x}$ is its mean, and $\sigma_{\bar{x}}$ the standard deviation \citep{Cheadle2003}. The features are normalized sample-wise one feature at a time: $\mathbf{x}$ is then a vector of length 120 containing all observations of a single feature in a sample, for example, all of the ask level 5 volumes. Separate normalization for different features is necessary due to the vastly different behaviors and scales between both different levels and volumes as well as their indicators. Including different indicators calculated from the limit order book, such as the price differences, allows for the preservation of information regarding the relations between different values, even after normalization.

The data are normalized sample-by sample due to the changes in price behavior that occur even during a single day. A relatively short normalization window is also needed to avoid larger scale price dependence. If, for example, the data were normalized over the full-time period, the main differences between prices in observations would come from the long time drift instead of the price changes in the recent past. As long-term changes are unlikely to be the main determining factor of jump occurrence in minute-level data, the normalization period should be short enough to avoid learning from them.

Additionally, in the used data, the most important factors seem to be changes that occur in the hours right before the jump. Changes within this timespan have also been noted for bigger jumps associated with company announcements, where changes in liquidity often start over an hour before the price jump \citep{Siikanen2017,siikanen2017drives}. The normalization done within the sample also requires a sufficiently big observation window, as it needs to be large enough to capture the element of change. There is also a fairly signficant chance that a jump has already occurred on the same day at the time another prediction is made, lessening the impact of price changes compared to the previous data.
Additionally, since all samples are of equal length, for the first two hours of the day, the window must include samples collected from the previous day.

\section{Neural Network Models}\label{SEC:NeuralNetworks}
Neural networks are learning systems that are modeled based on the structure of the human brain: large amounts of individual units, called neurons, process the information fed through the network. They then adjust their inner weights based on the information provided, making the system ``learn''. Methodwise, price jump prediction can be seen as a similar problem to mid-price prediction \citep{Kercheval2015a,ntakaris2017benchmark,tsantekidis2017using,tsantekidis2017forecasting,passalis2017time,tran2018temporal,sirignano2018universal}, although it has its own problems due to the small proportion of time-intervals with jumps versus without jumps. The methods used in this work are the standard MLP, LSTM, and convolutional networks, which are chosen due to their success in the prediction and classification of other time series \citet{yang2015deep,xingjian2015convolutional,greff2017lstm}. Moreover, a new network model is developed by combining convolutional and LSTM layers as well as the attention model proposed by \citet{Zhou2016}. The proposed convolutional Long Short-Term Memory Attention model (CNN-LSTM-Attention) aims to utilize LSTM for time series memory and CNN and the attention model for reducing the input size, increasing locality, and focusing on the most important features to improve prediction results.

\subsection{Multi-layer perceptron}
Perhaps the most common type of neural network is the MLP, which is a feed-forward neural network formed by layers of neurons stacked in a hierarchical manner. It receives the data vectors in the input layer, and then the information is propagated throughout the hidden layers, providing a response at the output layer. Each layer is formed by a set of neurons, each receiving an input from the neurons of the preceding layer, and provides a nonlinear response of the form
\begin{equation}
b_h = \theta_h\left(\sum^I_{i=1}{w_{ih}x_i}\right),
\end{equation}
where $I$ is the number of neurons in the previous layer, each providing an input $x_i$, and $w_{ij}$ is the weight connecting the $i$-th neuron in the preceding layer to the $j$-th neuron of the current layer. $\theta_{\cdot}$ is a nonlinear (piece-wise) differentiable function, which is used to nonlinearly scale the response of the neuron. The output neuron works exactly as the hidden layer neurons, although they may use a different activation (e.g., to lead to probability-like responses).

The optimal size of the hidden layer is defined by the data used, whereas the output layer size is defined by the number of output classes \citep{Graves2012,Haykin2004}. Multi-class classification is performed by following a competitive training approach, that is, the output neuron with the highest response indicates the predicted class label \citep{chollet2015keras}.

The training of a network consists of two phases, \textit{forward pass} and \textit{backward pass}. In forward pass, training vectors are introduced to the network and its responses are obtained. These responses are used in combination with the provided annotations (i.e., target vectors indicating the optimal response for each training vector) to define the network's error with respect to a loss function. This error is then used in the backward pass to update the parameters of the network. This is achieved by exploiting the (piece-wise) differentiable property of the neurons' activation functions, following a gradient descent learning approach called error backpropagation. We use an advanced version of this parameter update approach, called Adam \citet{Kingma2014}, which adaptively defines the hyper-parameters of each update step based on the input vectors. 

For classification problems and networks giving probability-like responses, the crossentropy loss function is commonly used. It determines the entropy between sets by measuring the average number of bits needed to identify an event drawn from a set. For discrete sets $p$ and $q$, where $p_i$ is the true label and $q_i$ is the current predicted value, binary crossentropy can be defined as
\begin{equation}
H(p, q) = - \sum_i{p_i log(q_i)}.
\end{equation}
It can be shown that when choosing between distributions $q$, which estimate the true distribution $p$, minimizing cross-entropy leads to choosing the best estimate by maximizing the overall entropy \citep{Shore1980}. Thus, it is a suitable loss function to be minimized, and often portrays the true loss better than simple error measures.

\subsection{Recurrent neural networks and Long Short-Term Memory}

In this paper, the Long Short-Term Memory (LSTM) model is used to accumulate features in time-domain and to simulate memory, by passing the previous signals through the same nodes. LSTM can be seen as a special case of recurrent neural networks (RNN) in which the connections between neurons allow directedly cyclical connections. In a basic recurrent network, neurons form connections inside the same layer, creating a net of one-way connections. In the simplest form, this means a standard neural network but with a feedback loop. The connections in the basic RNN are weighted as in a standard MLP. RNNs address the temporal relationships in their inputs by maintaining an internal state due to the recursive property, a quality especially suitable for time series data \citep{Giles2001}.

LSTM was first proposed by \citet{Hochreiter1997} and it was developed to combat the problem of keeping error signals in proportion when flowing backward in time (especially for long time dependencies) by making use of both short-term memory, based on the recurrent connections, and long-term memory, represented by the slowly changing weights. A constant error signal flow is ensured by connecting the neurons to themselves. LSTM introduced the concept of a memory cell to control the memory flow of a network. A memory cell is a singular neural unit with the addition of multiplicative input and output gates. These are created to protect the neuron from changes triggered by irrelevant inputs and to protect other units from the irrelevant information currently stored within the neuron. Each memory cell has a fixed self-connection and processes input from multiple input sources to create the output signals. Memory cells that share the same input and output gates form memory cell blocks \citep{Hochreiter1997}.

Training an LSTM network is done using a modified version of backpropagation, where a single step involves a forward pass and the update of all units through the computation of error signals for all weights, which are passed backwards in the network (backward pass). The activation of the input gate $y^{in}$ and output gate $y^{out}$ are defined as
\begin{equation}
y^{out_j}(t) = f_{out_j}(\sum_{m}w_{out_jm}y^m(t-1)),
\end{equation}
\begin{equation}
y^{in_j}(t) = f_{in_j}(\sum_{m}w_{in_jm}y^m(t-1)),
\end{equation}
where $j$ is the memory block index and $v$ is a cell inside the memory block $j$, so that $c_j^v$ marks the $v$-th cell of the $j$-th memory block and $w_{lm}$ is the weight for the connection between units $m$ and $l$. Input gates are defined as $in$ and output gates as $out$. The loop sums all the source units defined by the network. The function $f$ is a differentiable function for the gates, such as the logistic sigmoid
\begin{equation}
f(x) = \frac{1}{1 + e^{-x}},
 \end{equation}
where $x \in [0, 1]$.
The input is further squashed by a differentiable function $g(\cdot)$ \citep{Gers1999}.

\citet{Gers1999} further adds to the LSTM model by including an additional gate, the ``forget gate''. The forget gate allows the LSTM cell to reset itself at appropriate times, releasing resources to use. The LSTM layer outputs either a one-dimensional vector of activations for each feature or a two-dimensional structure with a value for each feature at each processed time step. With an LSTM layer connected to a dense layer, the former is needed, as the dense layer expects one-dimensional input. However, some models, such as the attention model proposed by \citet{Zhou2016}, require multi-dimensional LSTM output when applied to the LSTM layer, as  its purpose is to calculate a weighting value for each time step.

\subsection{Convolutional neural networks}
Convolutional neural networks (CNN) can be used to capture patterns in time and feature space. Convolution neurons combine information from neighboring observations in the feature and/or time dimensions and each neuron identifies different pattern in the input time-series.

CNNs mimic the way the visual system processes visual data. Specific neurons are only concerned with specific parts of the input, simultaneously making the position of specific features less relevant, as long as they are in a certain relation to the other features. Even though they were originally proposed for image recognition tasks, CNNs have found uses in speech classification and time series prediction tasks. The convolutional network combines the principles of the importance of locality in data points, shared weights between points, and possible subsampling.  \citep{LeCun1995} CNNs have been especially successful in the domain of image processing, providing, for example, a winning best entry in the popular ImageNet image classification challenge \citep{Krizhevsky} and the ImageNet feature localization challenge \citep{Sermanet2013}. In a CNN, the images are first normalized, resized, and approximately centered. After the input layer, each unit in a single layer receives inputs from a certain set of inputs in its neighborhood from the previous layer, making the receptive fields localized. This allows the extraction of certain local features, which can then be combined \citep{LeCun1995}.

\begin{figure}
\begin{center}
      \includegraphics[width=1.0\textwidth]{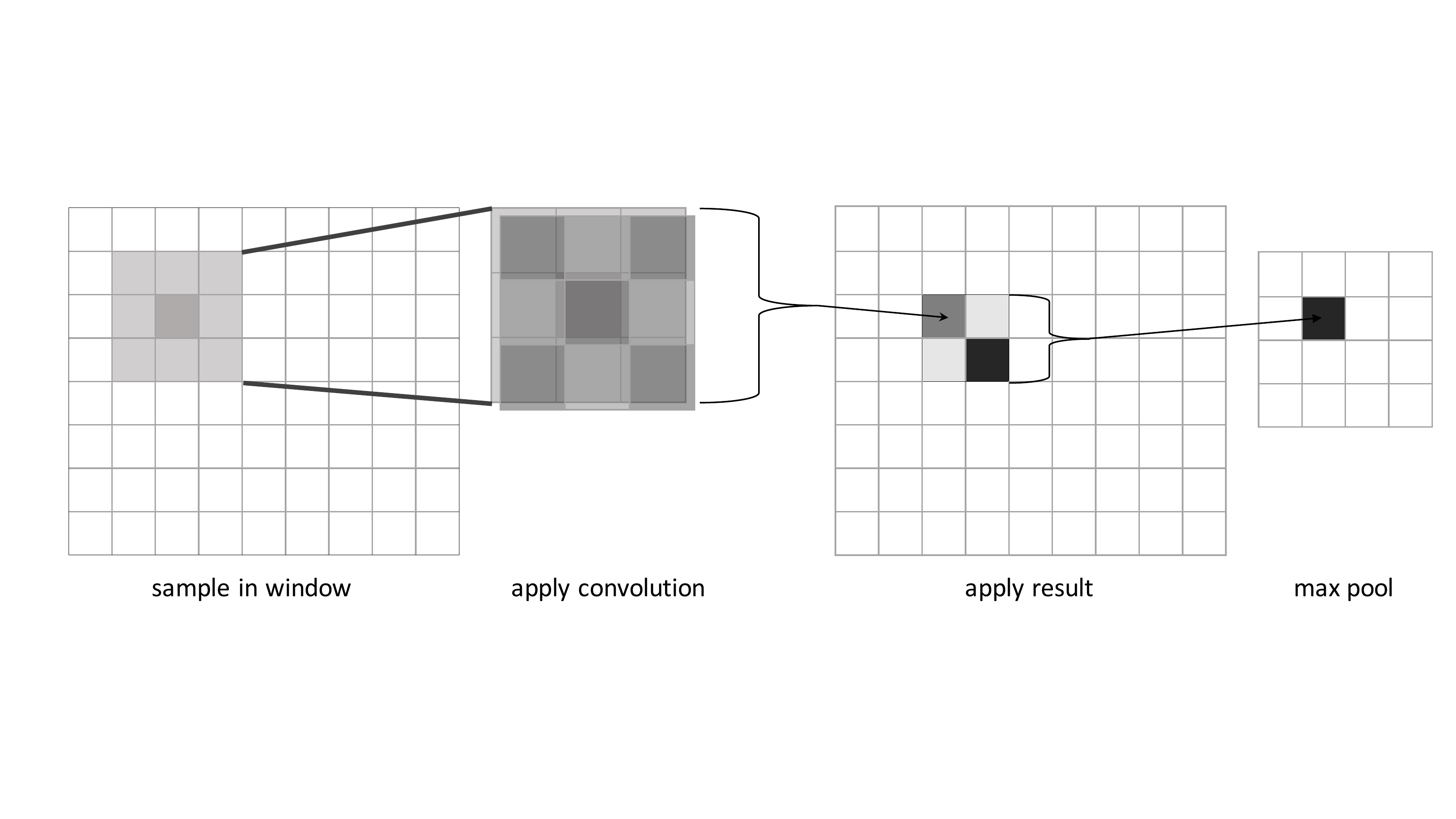}
\end{center}
	  \caption{\textbf{2d convolution with max pooling.} A single 2d CNN layer,
	  taking the convolution neighborhood, applying the
	  convolution kernel and reducing dimensionality with max pooling. (Adapted from \citep{Sermanet2013})}
	  \label{fig:cnn1} 
\end{figure}

Each convolutional layer is followed by an additional ``pooling layer'' to perform local averaging and/or subsampling. This reduces the resolution of the input at every step and reduces the network's sensitivity to shifts and distortions \citep{LeCun1995}. A simple CNN-pooling combination is shown in Figure \ref{fig:cnn1}. Pooling can also be done using the maximums of the input window, drawing attention to more pronounced features while reducing the resolution. This is called max pooling and is also often done between convolutions \citep{Scherer2010}. Convolutional and pooling layers are usually repeated until the feature maps convolute to a singular output for all possible classification results \citep{LeCun1995}, or they may be connected to regular dense (MLP) network layers to produce the final output \citep{Krizhevsky}.

\begin{figure}
\begin{center}
      \includegraphics[width=1.0\textwidth]{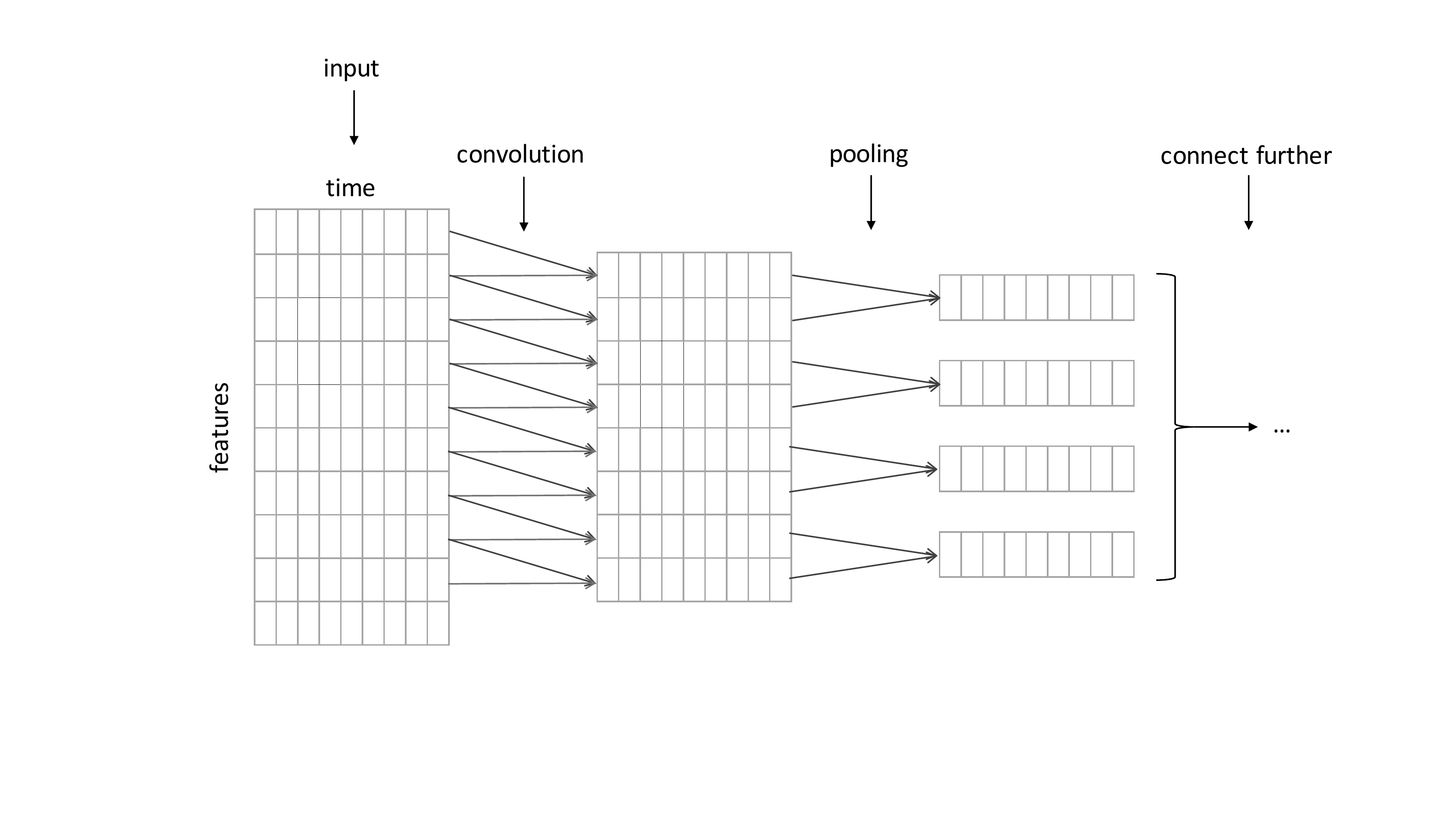}
\end{center}
          \caption{\textbf{1d convolution with pooling.} A single 1d CNN layer,
          convoluting in feature dimension, applying the
          convolution kernel and reducing dimensionality with
	  unspecified pooling. (Adapted from \citep{Hu2014})}
\label{fig:cnn2}
\end{figure}

Time series analysis with convolutional neural networks works much the same as in images, although the dimensionalities of the inputs are naturally different. Locality of the fields works well with time series, as the observations are dependent on time; the same observation can be followed by different results at different times, and the surroundings of the observation can be used to generate a better estimate \citep{Langkvist2014}. Convolutions can also be applied to one-dimensional time series data, allowing the convolution for both single- and multi-parameter problems \citep{DiPersio2016}. An example of feature-dimension time series convolution is presented in Figure \ref{fig:cnn2}.

\subsection{Dropout}
Dropout layers, first proposed by \citet{Hinton2012}, improve classification results by preventing complex co-adaptations of the training data. On each introduction of a training sample, hidden units are randomly omitted, according to a probability distribution, thus ``dropping out'' the unit activations from the information flow. As they may not be present, this means hidden units cannot rely on the presence of any other hidden unit at any time, making the network more robust as it cannot depend on any single passed value.

The probability of dropping out any one unit is predefined; \citet{Hinton2012} proposes a dropout threshold of 0.5. This means that generally only half of the units are present at any iteration of the training, and thus even if they fully (over)fit into a given training sample, the entire network will not. Dropout can be introduced with any connection, for example, between layers, or inside the recurrent connections of an LSTM layer.

\subsection{Attention model}
Attention is a mechanism that has been recently used in sentence classification, translation \citep{Bahdanau2014}, and generation \citep{Graves2013}. An attention mechanism generates an output by focusing on relevant elements of the input. That is, the attention model gives weights to the elements of the input sequence based on both the location and the contents of the sequence, supporting the possibility that observations at specific spots could have a greater importance in determining the results. Thus, the attention model could be used to weight different words in a sentence to find relations between them \citep{Zhou2016} or to weigh different time steps in a time series, for example, in speech recognition \citep{Chorowski2015}.

In this paper, we employ the attention layer proposed by \citet{Zhou2016} for sentence relation classification, with LOB data. Here, the steps are the timesteps of the LOB observations processed by the recurrent layer. In this model, the output representation $r$ is formed by a weighted sum of several output vectors:
\begin{equation}\nonumber
\begin{split}
M &= \text{tanh}(H)\\
\alpha &= \text{softmax}(w^TM)\\
r &= H\alpha^T,
\end{split}
\end{equation}
where $H$ is the attention layer input matrix consisting of the recurrent layer's output vectors $[h_1, h_2, ..., h_T]$, and $H \in R^{d^w \times L}$, where $d^w$ is the dimension of the observation vectors. $w$ is a trained parameter vector and $w^T$ its transpose; $L$ is the length of the sequence \citep{Zhou2016}. The $softmax$ is a normalized exponential function that squashes the inputs to output probability-like responses in the range $[0, 1]$:
\begin{equation}\nonumber
\text{softmax}(z_i) = \frac{e^{z_i}}{\sum_j e^{z_j}},
\end{equation}
where the activation is calculated in an element-wise manner \citep{Mikolov2015}. The final output of the attention layer is calculated from the representations with
\begin{equation}\nonumber
h^* = \text{tanh}(r).
\end{equation}
\citet{Zhou2016} also includes a softmax dense layer, which takes the attention output $h^*$ to calculate the final classification result \citep{Zhou2016}.

In this work, the attention layer is connected directly into the unconvoluted input, followed by the convolution and LSTM layers. Additionally, in place of time steps, the attention model is applied on the feature dimension. That is, all features are weighted, and the same weight for a single feature is repeated and thus applied to all of the time steps within the sample. This allows for selecting the features that are most relevant in any given sample.

\subsection{Implementation}
The neural networks were built using several Python libraries. The main library used was Keras, a high-level, open-source framework for building multilayer networks focused on enabling fast experimentation \citep{chollet2015keras}. Keras, however, does not provide the network structure but rather an interface for building it. Thus, TensorFlow, an implementation for executing different machine learning algorithms, was used as the Keras backend. Tensorflow is a flexible system, allowing the utilization of graphics processing units for speeding up the computation \citep{tensorflow2015-whitepaper}. The Keras' Model provides a simple framework to which layers can be added in a straightforward manner, and their connections to other layers can be specified. This allows the building of both simple sequential networks as well as more branching approaches. As Keras provides premade definitions for many different layer types,
experimenting with different configurations is fairly simple.

\begin{figure}
  \begin{center}
    \includegraphics[width=0.4\textwidth]{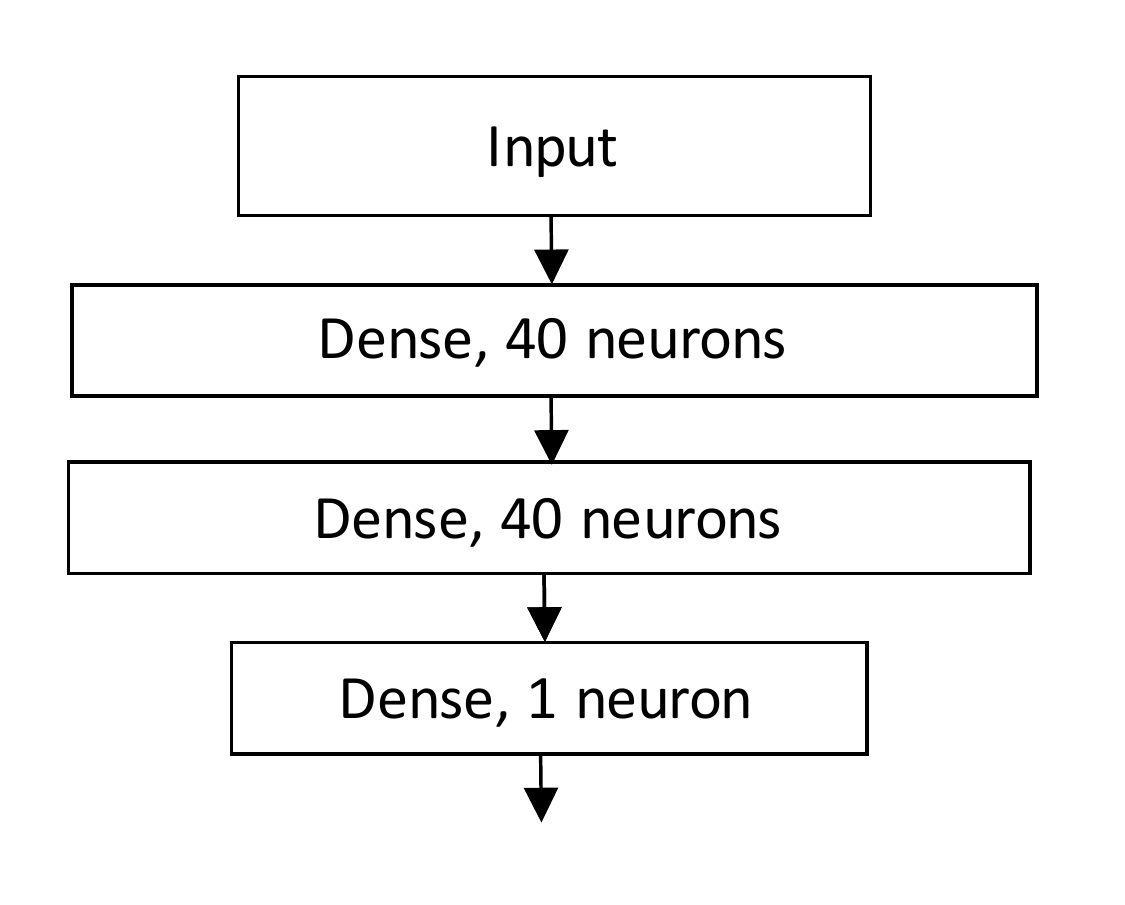}
  \end{center}
  \caption[Structure of the MLP model]{Layer structure of the MLP network used.}

  \label{fig:mlpill}
\end{figure}

\begin{figure}[!ht]
  \begin{center}
    \includegraphics[width=0.4\textwidth]{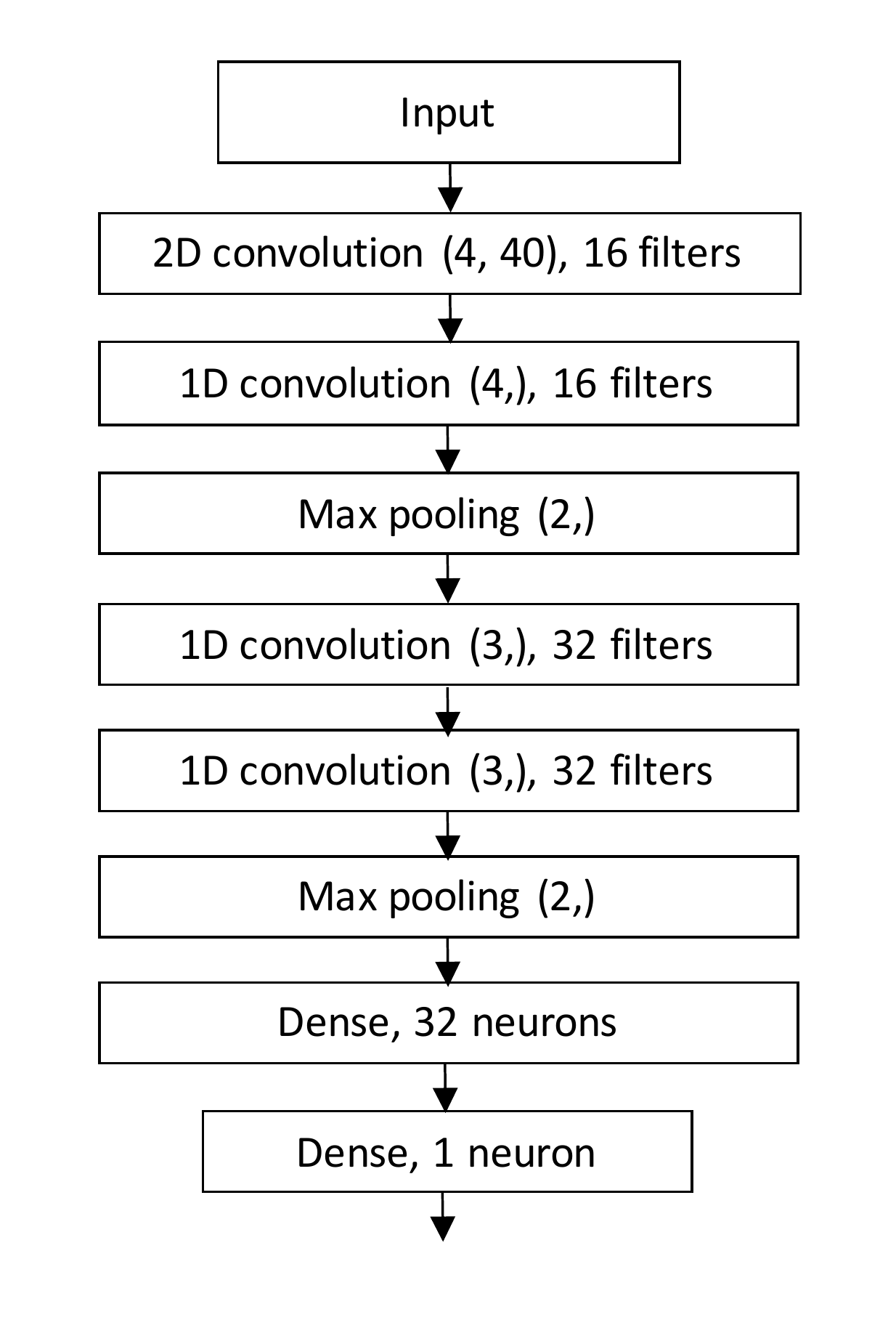}
  \end{center}
  \caption[Structure of the CNN model]{Layer structure of the convolutional network used.}

  \label{fig:cnnill}
\end{figure}

The MLP network consisted of two leaky ReLu layers of 40 neurons each. The MLP network structure is presented in Figure  \ref{fig:mlpill}. The CNN model for predicting stock price movements proposed by \citet{tsantekidis2017using} is illustrated in Figure \ref{fig:cnnill}. It consists of eight layers. The first layer is a 2D convolution with 16 filters of size (4,40), followed by an 1D convolution with 16 four-long filters and a max pool of two. This is followed by 2 additional 1D convolutions with 32 size 3 filters, and one additional size 2 max pooling layer. Furthermore, there are two fully connected dense layers, the first one with 32 neurons and the following one with 3 neurons. The output layer is modified to contain only a single output neuron to act as a two-class classifier. Additionally, while the network was designed to use only the 40 pure limit order book data features, it was modified in size to test it with the extra features used in this research. However, the original 40-feature network was selected for further analysis due to better results. The differences may have been due to the 2D convolution, which mixes features in both time and feature axes.

\begin{figure}[!ht]
  \begin{center}
    \includegraphics[width=0.4\textwidth]{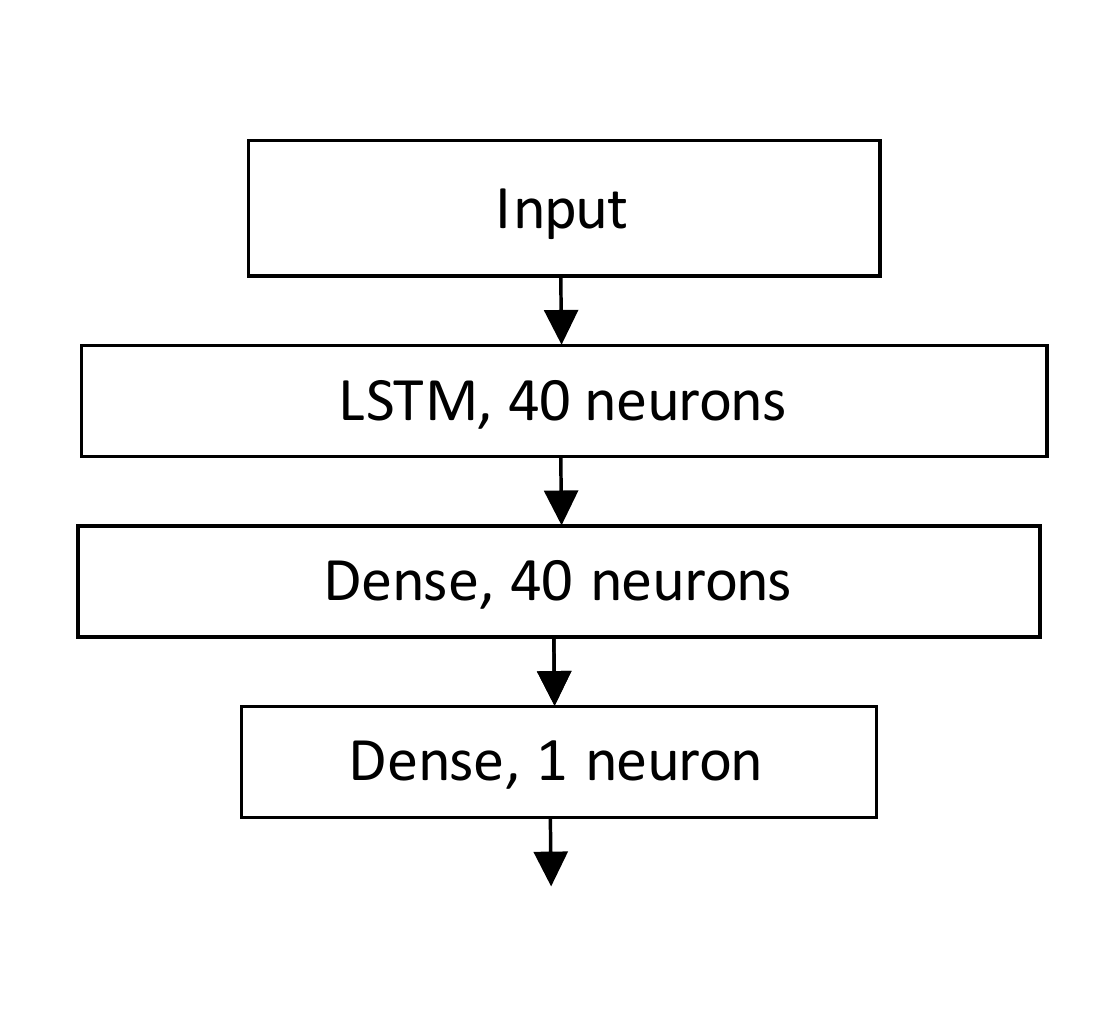}
  \end{center}
  \caption[Structure of the LSTM model]{Layer structure of the LSTM network used.}

  \label{fig:lstmill}
\end{figure}

Another network is the LSTM network for stock price prediction presented in \citet{tsantekidis2017using}. The LSTM network structure is shown in Figure \ref{fig:lstmill}. The network consists of an LSTM layer with 40 hidden neurons followed by a fully connected Leaky ReLu unit defined in \citet{Maas2013}.

The CNN-LSTM-Attention network is the most sophisticated model in this paper and it is designed to learn the most important patterns through feature and time domains for jump prediction and to optimally weight the different features to predict jumps. It is constructed as follows. The first layer connected after the input is the attention layer, composed of multiple Keras components: A regular dense layer with tanh activation is created with a weight for each time step, flattened to one dimension, to which softmax activation is further applied. This layer is repeated once for each step to apply the attention to full time steps. The dimensions are then switched to match the original input shape and merged together by multiplying the activations from the attention model and the input values from the original input layer. This gives each feature its own weight such that the same feature is weighted the same across all given time steps within a sample.

\begin{figure}[!ht]
  \begin{center}
    \includegraphics[width=0.4\textwidth]{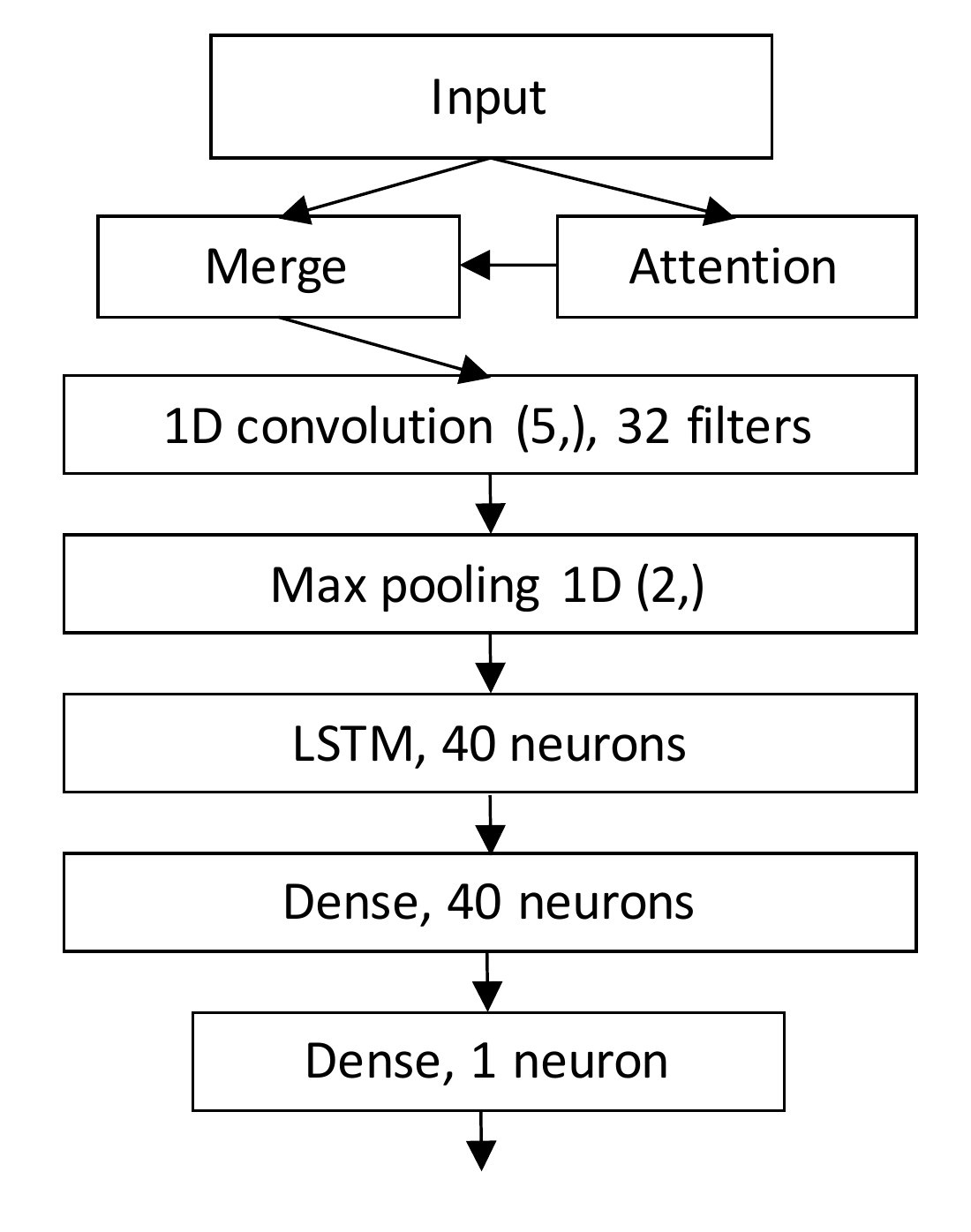}
  \end{center}
  \caption[Structure of the CNN--LSTM--Attention network]{Layer structure of the CNN-LSTM-Attention network. Additionally, the attention layer consists of repeated single neuron layers to apply activations on a time-step basis.}

  \label{fig:lstmcnnatt}
\end{figure}

The resulting attention mechanism output is a matrix of the original input size, which is passed forward to a 1D convolutional layer with 32 size 5 filters. The convolution output is further processed with a max pool of size 2, and the max pooled activations are passed to an LSTM layer with 40 relu neurons. The LSTM also includes a dropout of 0.5 both inside regular and recurrent connections. After the LSTM, there is a regular dense fully connected layer of the same size and, finally, the singular output neuron with sigmoid activation. This means that the output is a single value in the range $[0, 1]$, which is then rounded to obtain class prediction. The proposed network structure is illustrated in Figure \ref{fig:lstmcnnatt}.

\section{Results}\label{SEC:Results}

\subsection{Performance Measures}

The network performance was assessed with several metrics. The main target is F1 score, which is defined as the harmonic mean of precision and recall:
\begin{equation}
F1 = \frac{2}{\frac{1}{recall} + \frac{1}{precision}}
\end{equation}
Recall is defined as
\begin{equation}
recall = \frac{tp}{tp + fn}
\end{equation}
and precision as
\begin{equation}
precision = \frac{tp}{tp + fp}
\end{equation}
where $tp$ is true positives, the number of jump samples correctly classified as jumps; $fn$ is false negatives, jumps incorrectly classified as negative samples; and $fp$ is false positives, negative samples incorrectly classified as jumps. Thus, recall is the portion of jumps classified as jumps, and precision is the portion of real jump samples in samples classified as jumps \citep{Lipton}. High recall implies that a majority of jumps can be detected, whereas high precision means that jumps can be detected without also classifying many non-jump samples as jumps.

It should be noted that neither precision nor recall consider the number of true negatives. This also makes F1 independent of the ratio of accurately classified negatives, and instead focuses heavily on correctly classifying the positives. Thus, F1 provides a measure that is both non-linear and non-symmetric—F1 is commonly used in cases where the positive class is rare compared to the negatives \citep{Lipton}. As the portion of jumps in the data is very small, F1 is a suitable metric. Successfully predicting no jump is also a less informative result than successfully predicting one given a relatively small number of false positives, as very good accuracy could be achieved by just always predicting that there will be no jump. Measuring the results with F1 also avoids this scenario, as correct negatives do not affect the score.

Another used metric is Cohen's Kappa:
\begin{equation}
\kappa = \frac{p_o - p_c}{1 - p_c},
\end{equation}
for which $p_o$ and $p_c$ can be calculated from the confusion matrix \citep{Cohen1960}. As the Kappa also takes into account the agreement by chance, it can be seen as a more robust measurement for agreement. There is no single interpretation for what can be considered a good Kappa value, and thus it depends on the type of the problem analyzed.  \citet[p. 604]{Fleiss2003} presents intervals such that values greater than 0.75 signify excellent agreement, values above 0.40 a fair or good agreement, and values under 0.40 a poor agreement. Conversely, \citet{Landis1977} suggest that values of 0.21-0.40 are already fair, 0.41-0.60 are moderate, 0.61-0.80 are substantial, and values above that are almost perfect.

\subsection{Main results}

Performance of the networks used is presented by averaging the scores across all stocks and sets in Table \ref{table:results_avg}. Because of the unbalanced data, we consider F1 as the most appropriate performance measure, and therefore F1 values are in a bold font in the table. The table shows that of all the tested networks, the CNN--LSTM--Attention model achieves the highest average F1 for all samples (around 0.72). The second best network is the pure LSTM (0.69), followed by the CNN (0.66) and the CNN-LSTM-$v_{10}$, where no information other than the time of day (feature $v_{10}$) is used. Finally, the MLP achieved an average F1 of 0.53. All of the models clearly outperform a random classifier, for which F1 is 0.32.

Additionally, when comparing average F1s by stocks, according to Table \ref{table:results_overview}, the CNN-LSTM-Attention model again performs the best. With all of the tested network models, the resulting F1s are above those of a random classifier. MLP is somewhat worse, possibly due to being unable to deal with fairly large time series input data without overlearning, but it still clearly better than random choice. The scores implies that at least a part of the jumps in the data are predictable with a reasonable level of confidence. Additionally, when interpreting the Kappa scores, the scores of both LSTM models can be seen as at least good if not very good. Also the CNN-LSTM-$v_{10}$ works surprisingly well, given that jumps are predicted using information on the time of day only. Therefore, LOB data can be said to be useful for predicting price jumps, especially for Apple, Facebook, and Microsoft, but sometimes it gives a rather marginal advantage (see Google and Intel in Table \ref{table:results_overview}).

\begin{table}[!ht]
\centering
\bgroup
\def\arraystretch{1.2}
 \begin{tabular}{|c|cccc|}
 \hline
 & Precision & Recall & \textbf{F1 }& Cohen's Kappa \\
 \hline
CNN-LSTM-A & 0.66 & \underline{0.80} & \textbf{\underline{0.72}} & \underline{0.62} \\
LSTM & 0.73 & 0.66 & \textbf{0.69 }& 0.60 \\
CNN & 0.66 & 0.66 & \textbf{0.66 }& 0.55 \\
MLP & \underline{0.78} & 0.41 & \textbf{0.53} & 0.44 \\
Random & 0.24 & 0.50 & \textbf{0.32 }& 0.00 \\
CNN-LSTM-$v_{10}$ & 0.57 & 0.79 & \textbf{0.66} & 0.54 \\
 \hline
\end{tabular}
\egroup
\caption[Network performance averages]{Overall precision, recall, F1, and Cohen's Kappa scores for all four networks and a random classifier. As F1 is the most appropriated performance measure because of unbalanced data, it is emphasized in a bold font in the table. CNN-LSTM-$v_{10}$ denotes CNN-LSTM using no other information but the time of day (feature $v_{10}$).
}
\label{table:results_avg}
\end{table}

\begin{table}[!ht]
\centering
\scalebox{0.85}{
\bgroup
\def\arraystretch{1.2}

 \begin{tabular}{|c|c|cccc|}
 \hline

 && Precision & Recall & \textbf{F1} & Cohen's Kappa \\
 \hline
AAPL & CNN-LSTM-A & 0.67 & \underline{0.74} & \textbf{\underline{0.71}} & \underline{0.61}\\
  & LSTM & 0.70 & 0.54 & \textbf{0.61} & 0.50\\
  & CNN & 0.62 & 0.57 & \textbf{0.59}& 0.47\\
  & MLP & \underline{0.73} & 0.35 & \textbf{0.47} & 0.37\\
  & Random & 0.24 & 0.50 &\textbf{0.33} & 0.00 \\
  & CNN-LSTM-$v_{10}$ & 0.65 & 0.72 & \textbf{0.68} & 0.58 \\
  \hline
 FB & CNN-LSTM-A & 0.67 & 0.81 & \textbf{\underline{0.73} }& \underline{0.63}\\
  & LSTM & 0.73 & 0.67 & \textbf{0.70} & 0.60\\
  & CNN & 0.69 & 0.70 & \textbf{0.69} & 0.59\\
  & MLP & \underline{0.80} & 0.41 & \textbf{0.54} & 0.45\\
  & Random & 0.25 & 0.50 & \textbf{0.33} & 0.00 \\
  & CNN-LSTM-$v_{10}$ & 0.48 & \underline{0.90} & \textbf{0.62} & 0.44 \\
  \hline
 GOOG & CNN-LSTM-A & 0.60 & \underline{0.83} & \textbf{\underline{0.69} }& \underline{0.59}\\
  & LSTM & 0.69 & 0.64 & \textbf{0.66} & 0.58\\
  & CNN & 0.62 & 0.77 & \textbf{\underline{0.69}}& \underline{0.59}\\
  & MLP & \underline{0.75} & 0.36 & \textbf{0.48} & 0.40\\
  & Random & 0.21 & 0.50 & \textbf{0.30}  & 0.00 \\
  & CNN-LSTM-$v_{10}$ & 0.57 & 0.82 & \textbf{0.68} & 0.57 \\
  \hline
 MSFT & CNN-LSTM--A & 0.62 & \underline{0.77} &\textbf{\underline{0.69} }& 0.59\\
  & LSTM & 0.72 & 0.65 & \textbf{0.68} & \underline{0.60}\\
  & CNN & 0.66 & 0.66 & \textbf{0.66} & 0.56\\
  & MLP & \underline{0.76} & 0.42 &\textbf{0.54 }& 0.46\\
  & Random & 0.22 & 0.50 & \textbf{0.30 }& 0.00  \\
  & CNN-LSTM-$v_{10}$& 0.49 & 0.70 & \textbf{0.58} & 0.44 \\

  \hline
 INTC & CNN-LSTM-A & 0.72 & \underline{0.85} & \textbf{\underline{0.78}}& \underline{0.69}\\
  & LSTM & 0.77 & 0.77 & \textbf{0.77} & \underline{0.69}\\
  & CNN & 0.71 & 0.63 & \textbf{0.67} & 0.56\\
  & MLP & \underline{0.84} & 0.48 & \textbf{0.61} & 0.52\\
  & Random & 0.26 & 0.50 & \textbf{0.34} & 0.00 \\
  & CNN-LSTM-$v_{10}$ & 0.72 & 0.82 & \textbf{0.77} & 0.68 \\
 \hline
\end{tabular}
\egroup
}
\caption[Network performance overview by stock]{Overall precision, recall, F1, and Cohen's Kappa scores by network and stock for all four networks tested. The CNN-LSTM-Attention network has the best total average Recall, F1, and Kappa scores. The MLP has a higher precision with the cost of greatly reduced recall--the given class labels are skewed towards negative. Random is a fully random classifier that is included as a benchmark. By definition, random recall is 0.50 and Kappa is 0. As F1 is the most appropriated performance measure because of unbalanced data, it is emphasized in a bold font in the table. CNN-LSTM-$v_{10}$ denotes CNN-LSTM using no other information other than the time of day (feature $v_{10}$).
}
\label{table:results_overview}
\end{table}

According to Tables \ref{table:results_avg} and \ref{table:results_overview}, the most promising model is CNN-LSTM-Attention, and hence we zoom into the F1 scores of CNN-LSTM-Attention model over the stocks and data sets in Table \ref{table:f1_CNN_LSTM_Attention}. The average F1 for all CNN-LSTM-Attention model sets is 0.71, with some variation between both different time periods and different stocks. The variation is most likely due to different stock price and jump dynamics for different securities.\footnote{In Table \ref{table:f1_CNN_LSTM_Attention}. the scores are first calculated individually for each set and stock and then averaged over the sets. As the number of samples is not constant across the sets, the average of individually calculated set scores is not exactly equal to the score calculated directly across all samples. Because of this, there are minor deviations between the values presented in Tables \ref{table:results_avg}, \ref{table:results_overview} and \ref{table:f1_CNN_LSTM_Attention}. }  As true negatives are not accounted for in the calculation of F1, and having less jumps to detect directly lowers precision if the ratio of detected jumps remains the same, fewer jumps to detect tends to cause a lower score unless the amount of jumps that are easily predictable remains the same. Intel (INTC) jumps were predicted the best (0.78 F1) while Microsoft had the worst score, with a difference of around ten percentage points. The corresponding results for other models, the LSTM model proposed in \citet{tsantekidis2017forecasting}, the convolutional model proposed in \citet{tsantekidis2017using}, and a regular two-layer MLP network, are provided in the Appendix (Tables \ref{table:f1_lstm}, \ref{table:f1_cnn}, \ref{table:f1_mlp}).

\begin{table}[ht]
\centering
\bgroup
\def\arraystretch{1.2}

 \begin{tabular}{|c|ccccc|c|}
 \hline

Set/stock & AAPL & FB & GOOG & MSFT & INTC & Average  \\
 \hline
Set 1 & 0.76 & 0.63 & 0.68 & 0.68 & 0.74 & 0.70 \\
Set 2 & 0.70 & 0.67 & 0.47 & 0.72 & 0.84 & 0.68 \\
Set 3 & 0.66 & 0.72 & 0.69 & 0.68 & 0.74 & 0.70 \\
Set 4 & 0.65 & 0.85 & 0.69 & 0.68 & 0.81 & 0.73 \\
Set 5 & 0.75 & 0.81 & 0.78 & 0.66 & 0.79 & 0.76 \\
Set 6 & 0.72 & 0.75 & 0.77 & 0.76 & 0.74 & 0.75 \\
Set 7 & 0.70 & 0.69 & 0.74 & 0.53 & 0.77 & 0.68 \\
\hline
Average & 0.70 & 0.73 & 0.69 & 0.67 & 0.78 & 0.71 \\
 \hline

\end{tabular}
\egroup
\caption{F1 scores by set and by stock for CNN-LSTM-Attention network.}
\label{table:f1_CNN_LSTM_Attention}
\end{table}

\subsection{Attention layer}

Even though the attention model was originally proposed in \citet{Zhou2016} for use in the time dimension, it can also be used to highlight important features. That is, instead of focusing on time steps where interesting things occur, the network focuses on features that are particularly interesting at that time point. This was suitable due to the large number of features used.

Four samples were selected for qualitative analysis of the attention mechanism through the activations of the layer created for a given sample: one true positive and true negative as well as one false positive and false negative. As expected, all of these samples placed some importance on the time feature $v_{10}$ (see Table \ref{table:Features}), as its feasibility had already been studied while developing the network. The attention was checked by comparing the attention activations from the unmerged layer. Other than that, the attention was given to quite different features between the four samples.

Both true and false positives included some volumes from the top levels of the order book from the basic set ($v_1$) of the features. Still, neither included any price information from the books in the attention, even though the quantities were regarded as important. Both also included several derivatives from the set of features $v_6$; the true positive had only quantities from both sides, whereas the false positive also included ask prices. Interestingly, the real positive sample also focused on a single, level 4 ask price difference from feature set $v_3$, none other of which were featured in any other inspected samples.

The negative samples were fairly different from the positives, although both had focused on the time feature as well as some derivative from feature set $v_6$. However, the derivatives of the negative samples are purely quantity derivatives for ask or bid volumes. The false negative and true negative were also fairly similar to each other in regards to attention, which makes sense since the end classification result was the same. Both focused on the basic set for both ask and bid as well as quantity and price. They also focused on several mid prices, which was not the case with either positive sample. Interestingly, in addition to these, the true negative sample had also selected cancel intensities for both ask and bid.

For all of the inspected samples, the focus was clearly on several specific feature sets, with some features not included in the attentions at all.
However, ask and bid values were fairly equally present as well as values from across the 10 levels of the order books. Quantity-dependent values were regarded as especially important in multiple feature sets, which indicates the relation between liquidity and jumps \cite[see][for the relation between order book liquidity and news announcements]{siikanen2017drives,Siikanen2017}. Still, the full test set was not inspected, so even though this implies some relevance for the values to which attention was paid, the remaining ones would need to be thoroughly inspected to draw conclusions regarding their usefulness in jump prediction.

\subsection{Prediction of the direction of jumps}

This study focused on a two-class prediction problem, where the class information related to whether or not the jump statistic would exceed a certain threshold within the next minute independently of the direction of the jump. However, the problem can also be formulated as a three-class classification problem, where $c_0$ is no jump, $c_1$ is an upwards jump, and $c_2$ is a downwards jump. This was done with the same model and the same input data, but the output layer was changed to a three-neuron block. Still, this formulation proved much more challenging than only predicting the arrical of a jump, and the results did not significantly differ from random selection with regard to the jump direction. Results for differentiating between the jump classes $c_1$ and $c_2$ with the CNN-LSTM-Attention network are presented in Table \ref{table:B} in Appendix.

The better predictability of the arrival of a jump (than the sign of a jump) is consistent with the literature, which has provided evidence about the relation between jumps and news announcements and analysts' recommendations, whose arrival time, but not direction, is often predictable.  \cite{Lee2012,bradley2014analysts} This result is also consistent with the definition of the stock price model with jumps: the jump term consists of two separate parts, the counting process of the occurrence of a jump and the jump size, which is  independent and identically distributed
. Thus, it is possible that the counting process is predictable, while the size and thus the direction of the jump is not. However, this study did not focus on such a prediction case, and so it may be possible to also predict the direction of the jump with some accuracy, albeit likely much lower than the occurrence of the jump, which may even be known beforehand due to a prescheduled release of information regarding the stock.

Another possible direction for jump prediction is to not only predict jumps but to predict the jump statistic for each time step. This means that the prediction would be a regression problem in which the jumps are only implied by the statistics that are above the threshold by interpreting the output of the network. However, this method requires more accurate predictions of price movements considering the non-jump price process $\sigma(t)dW(t)$, and thus it also requires the prediction of the stock price process. Of course, the task can be made easier by using only the absolute value of the jump statistic, disregarding the direction of the movement. Even so, this task was much more difficult than simple jump prediction, at least with the tested methods. This is most likely due to the combination of more exact prediction requirements of a regression task, the limitations of the chosen methods, and the need to predict the magnitude of the normal stock price process to get the correct statistics in the samples without jumps. This question will be addressed further in our future research.

\section{Conclusions}\label{SEC:Conclusions}

In this work, a new CNN-LSTM-Attention model was developed to predict jumps from LOB data. The problem was also tested on several existing neural network models for stock price prediction. The networks were both trained and tested for five separate stocks for a total of 360 days. The developed CNN-LSTM-Attention model performed the best in regard to F1 for all stocks and, on average, also Cohen's Kappa. Overall, the F1 scores achieved with the other tested networks provide evidence that the use of limit order book data was found to improve the performance of the proposed model in jump prediction, either clearly or marginally, depending on the underlying stock. This suggests that path-dependence in limit order book markets is a stock specific feature. Predicting the direction of such jumps is much more difficult as is the jump statistic in general, which in line with the efficient market hypothesis.

Moreover, we find that the proposed approach with an attention mechanism outperforms the multi-layer perceptron network as well as the convolutional neural network and Long Short-Term memory model. 

This research focused mainly on predicting the existence of a jump in the near future. Predicting the jump statistic---rather than if the statistics exceeded a threshold or not---would be especially interesting even outside the context of the jumps, as it is in essence a statistic of future returns of the stock. Thus, it could also be of use in those contexts, even if the absolute value is used and the stock price direction is thus not considered. Additionally, the definition of jump itself includes some randomness due to the unbounded nature of the normal distribution, which would be eliminated from the learning process if it only considered the jump statistic that is directly based on the stock price and thus does not include uncertainty in its interpretation. This study also focused on several known network types, LSTM and CNN, to provide a basis for the model, and it might be beneficial to apply other types of classifiers to this type of a problem. This problem about predicting the statistics as a non-linear regression problem will be addressed in our future research.

Another area of interest is the feature attention model used in this study, as such models can also be also as indicators as to which types of measures can be used to predict price jumps in general and which features of the book affect the formation of the jump. However, the attentions were only studied based on several sample series, and so more in-depth research on these features would be particularly interesting. The results of the attention study could also be used to further develop the input data of the predictor network by expanding on the features that the network considers as relevant to classification. At the same time, performance in terms of results and computational complexity could be improved by leaving out features that do not yield significant
weights for any correctly classified samples.

\bibliographystyle{rQUF}
\bibliography{referenceFile}

\section*{Appendix}

\begin{table}[h]
\centering
 \begin{tabular}{|c|ccccc|c|}
 \hline
Set/stock & AAPL & FB & GOOG & MSFT & INTC & Average\\
\hline
Set 1 & 0.71 & 0.59 & 0.61 & 0.71 & 0.77 & 0.68\\
Set 2 & 0.36 & 0.63 & 0.56 & 0.82 & 0.83 & 0.64\\
Set 3 & 0.64 & 0.74 & 0.61 & 0.73 & 0.72 & 0.69\\
Set 4 & 0.59 & 0.72 & 0.68 & 0.64 & 0.81 & 0.69\\
Set 5 & 0.69 & 0.77 & 0.74 & 0.56 & 0.75 & 0.70\\
Set 6 & 0.67 & 0.71 & 0.74 & 0.65 & 0.79 & 0.71\\
Set 7 & 0.60 & 0.70 & 0.72 & 0.52 & 0.75 & 0.66\\
\hline
Average & 0.61 & 0.70 & 0.67 & 0.66 & 0.77 & 0.68\\
 \hline

\end{tabular}
\caption{F1 scores by set and by stock for LSTM network.}
\label{table:f1_lstm}
\end{table}

\begin{table}[h]
\centering
 \begin{tabular}{|c|ccccc|c|}
 \hline
Set/stock & AAPL & FB & GOOG & MSFT & INTC & Average\\
\hline
Set 1 & 0.57 & 0.65 & 0.69 & 0.68 & 0.74 & 0.67\\
Set 2 & 0.48 & 0.57 & 0.56 & 0.78 & 0.76 & 0.63\\
Set 3 & 0.60 & 0.69 & 0.62 & 0.69 & 0.68 & 0.66\\
Set 4 & 0.60 & 0.82 & 0.68 & 0.70 & 0.38 & 0.64\\
Set 5 & 0.59 & 0.77 & 0.77 & 0.55 & 0.62 & 0.66\\
Set 6 & 0.73 & 0.72 & 0.78 & 0.46 & 0.74 & 0.69\\
Set 7 & 0.63 & 0.47 & 0.69 & 0.62 & 0.61 & 0.60\\
\hline
Average & 0.60 & 0.67 & 0.68 & 0.64 & 0.64 & 0.65\\
 \hline

\end{tabular}
\caption{F1 scores by set and by stock for CNN network.}
\label{table:f1_cnn}
\end{table}

\begin{table}[h]
\centering
 \begin{tabular}{|c|ccccc|c|}
 \hline
Set/stock & AAPL & FB & GOOG & MSFT & INTC & Average\\
\hline
Set 1 & 0.50 & 0.51 & 0.50 & 0.23 & 0.47 & 0.44 \\
Set 2 & 0.21 & 0.19 & 0.46 & 0.79 & 0.69 & 0.47 \\
Set 3 & 0.43 & 0.36 & 0.49 & 0.72 & 0.60 & 0.52 \\
Set 4 & 0.45 & 0.63 & 0.63 & 0.40 & 0.65 & 0.55 \\
Set 5 & 0.56 & 0.78 & 0.47 & 0.45 & 0.59 & 0.57 \\
Set 6 & 0.63 & 0.65 & 0.18 & 0.43 & 0.69 & 0.52 \\
Set 7 & 0.57 & 0.66 & 0.54 & 0.57 & 0.62 & 0.59 \\
\hline
Average & 0.48 & 0.54 & 0.47 & 0.51 & 0.62 & 0.52 \\
 \hline
\end{tabular}
\caption{F1 scores by set and by stock for MLP network.}
\label{table:f1_mlp}
\end{table}

\begin{table}[h]
\centering
 \begin{tabular}{|c|ccccc|c|}
 \hline
Set/stock & AAPL & FB & GOOG & MSFT & INTC & Average\\
\hline
Set 1 & 0.68 & 0.60 & 0.64 & 0.56 & 0.73 & 0.64 \\
Set 2 & 0.59 & 0.54 & 0.35 & 0.26 & 0.87 & 0.52 \\
Set 3 & 0.71 & 0.57 & 0.72 & 0.76 & 0.71 & 0.70 \\
Set 4 & 0.64 & 0.61 & 0.61 & 0.71 & 0.75 & 0.66 \\
Set 5 & 0.75 & 0.74 & 0.76 & 0.70 & 0.79 & 0.75 \\
Set 6 & 0.76 & 0.76 & 0.82 & 0.77 & 0.78 & 0.78 \\
Set 7 & 0.69 & 0.66 & 0.71 & 0.51 & 0.71 & 0.66 \\
\hline
Average & 0.69 & 0.64 & 0.66 & 0.61 & 0.76 & 0.67 \\
\hline
\end{tabular}
\caption{F1 scores by set and by stock for $v_{10}$ network.}
\label{table:f1_v10}
\end{table}

\begin{table}[h]
\centering
\begin{tabular}{|c|c|cccc|}
\hline
Set & Method & Precision & Recall & Cohen's Kappa & \textbf{F1} \\
\hline
AAPL & CNN-LSTM-Att & \underline{0.45} & 0.49 & \underline{0.08} & \textbf{\underline{0.47} }\\
 & random & 0.41 & \underline{0.50} & 0.00 & \textbf{0.45 }\\
\hline
FB & CNN-LSTM-Att & \underline{0.58} & \underline{0.64} & \underline{0.08} & \textbf{\underline{0.61} }\\
 & random & 0.55 & 0.50 & 0.00 & \textbf{0.52 }\\
\hline
GOOG & CNN-LSTM-Att & 0.45 & \underline{0.59} & 0.00 & \textbf{\underline{0.51} }\\
 & random & 0.45 & 0.50 & 0.00 & \textbf{0.47 }\\
\hline
MSFT & CNN-LSTM-Att & \underline{0.54} & 0.50 & \underline{0.07} & \textbf{\underline{0.52} }\\
 & random & 0.50 & 0.50 & 0.00 & \textbf{0.50 }\\
\hline
INTC & CNN-LSTM-Att & 0.50 & \underline{0.51} & -0.01 & \textbf{\underline{0.51} }\\
 & random & \underline{0.51} & 0.50 & \underline{0.00} & \textbf{0.50 }\\
\hline
Average & CNN-LSTM-Att & \underline{0.51} & \underline{0.55} & \underline{0.05} & \textbf{\underline{0.53}} \\
 & random & 0.48 & 0.50 & 0.00 & \textbf{0.49 }\\
\hline

\end{tabular}

\caption[Predicting jump direction]{Results of the CNN-LSTM-Attention model for separated jump direction prediction, with upward jumps as positive samples and downward jumps as negative samples. The number of samples in the two classes is not equal, but the dominant direction depends on the set. The network is slightly better than a random classifier on average, although not consistently, and the results vary due to changing proportions of jump directions between different stocks. It should be noted that, in some sets, the number of upward jumps is large enough to make always choosing positive the best choice, even according to F1, as the changing imbalance between class sizes means that this option cannot be eliminated with the choice of positive and negative labels.}
\label{table:B}
\end{table}

\end{document}